 \documentclass[12pt]{article}
 \usepackage{amssymb}
 \usepackage{latexsym}
 \usepackage{graphicx}
 
 \renewcommand{\thefootnote}{\fnsymbol{footnote}}

 \def\appendix#1{
   \addtocounter{section}{1}
   \setcounter{equation}{0}
   \renewcommand{\thesection}{\Alph{section}}
   \section*{Appendix}
   \addcontentsline{toc}{section}{Appendix \thesection\ \ \ #1}
   }

 \newcommand{\newsection}{    
 \setcounter{equation}{0}
 \section}

 \newcommand{\eq}[1]{Eq.~(\ref{#1})}

 \def\bea{\begin{eqnarray}}
 \def\eea{\end{eqnarray}}

 \def\be{\begin{equation}}
 \def\ee{\end{equation}}
 
 \def \bi{\bibitem}

 \def\e9{\mbox{${E_9}$}}

 \def\id{\protect{{1 \kern-.28em {\rm l}}}}

\begin{document}

 \setcounter{page}{1}
 \renewcommand{\thefootnote}{\arabic{footnote}}
 \setcounter{footnote}{0}

 \begin{titlepage}
 \begin{flushright}
 CU-TP-1043\\
 \end{flushright}
 \vspace{1cm}

 \begin{center}
 {\LARGE Cosmological String Gas on Orbifolds}

 \vspace{1.1cm}
 {\large{Richard Easther$^*$\footnote{
 E-mail: easther@physics.columbia.edu}, Brian R. Greene$^{*\S}$\footnote{
 E-mail: greene@physics.columbia.edu}, Mark G. Jackson$^\dagger$\footnote{
 E-mail: markj@physics.columbia.edu}} }\\

 \vspace{18pt}

 {\it $^*$Institute for Strings, Cosmology and Astroparticle Physics}

 {\it $^\S$Department of Mathematics}

 {\it $^\dagger$Department of Physics}

 {\it Columbia University}

 {\it New York City, NY 10027}
 \\
 \end{center}
 \vskip 0.6 cm

\begin{abstract} It has long been known that strings wound around 
incontractible cycles can play a vital role in cosmology.  In particular, 
in a spacetime with toroidal spatial
hypersurfaces, the dynamics of the winding modes may help yield three 
large spatial dimensions. However, toroidal compactifications are 
phenomenologically unrealistic. In this paper we therefore take a first 
step toward extending these cosmological considerations to $D$-dimensional 
toroidal orbifolds.  We use numerical simulation to study the timescales over 
which ``pseudo-wound" strings unwind on these orbifolds with trivial 
fundamental group.  We show that pseudo-wound strings can persist for
many ``Hubble times'' in some of these spaces, suggesting that they
may affect the dynamics in the same way as genuinely wound strings. We 
also outline some possible extensions that include higher-dimensional wrapped 
branes.
\end{abstract}

\end{titlepage}
\newsection{Introduction}
 
A great deal of effort has been expended on studying compactifications
of string theory, that is, geometrical backgrounds in which four spacetime
dimensions are assumed to be large while the others are
unobservably small. In order to develop a complete and viable physical
model, though, we must not only describe the nature of the
compactification, but also understand the dynamical processes which
produced it. While a compactification can be, and usually is, imposed
by hand, doing so goes against the grain of the superstring program.

We are already familiar with a number of fundamental symmetries which
are broken as the universe evolves.  A common and reasonable
assumption that the division between large and small spatial
dimensions is one of the asymmetries arising naturally during the
course of cosmological evolution. For example, in the very early
universe all of the dimensions may have been compact with small radius
and some dynamical process may have subsequently driven significant
expansion in only three of the spatial dimensions. Some time ago,
Brandenberger and Vafa \cite{bv} proposed just such a mechanism for a
universe in which all spatial dimensions were toroidal. Here, wound
strings impede cosmological expansion if they fall out of thermal
equilibrium. Since two-dimensional string worldsheets generically fail
to intersect in more than four spacetime dimensions, the interactions
required for thermal equilibrium are not robust in more than four
spacetime dimensions, giving a qualitative explanation of why spatial
regions with more than three dimensions do not expand. This is an
attractive proposal, since it uses the simple and intrinsically
stringy winding modes. Moreover, aspects of the model were verified
\cite{sak} using a numerical algorithm originally devised by Smith and
Vilenkin \cite{vil} and this line of argument was recently extended to
include a spectrum of branes \cite{loit}.

Unfortunately, toroidal compactifications do not lead to realistic
particle physics models whereas toroidal orbifolds or Calabi-Yau
compactifications of string theory (or perhaps $G_2$
compactifications of M-theory) have a better chance of yielding the
particle phenomenology we observe. Obviously, it is worth rethinking
the Brandenberger-Vafa mechanism in these contexts. In this paper we
take a preliminary step in this direction.

A crucial ingredient of the Brandenberger-Vafa mechanism is that tori
have nontrivial cycles in every dimension, and each of these cycles is
a member of a family that moves throughout the torus.  This is not
true of more general compactifications. In particular, many toroidal
orbifolds and Calabi-Yau spaces have trivial fundamental groups and
thus no notion of wound strings.  Without incontractible 1-cycles it
would seem that one cannot expect an analog of the Brandenberger-Vafa
mechanism to hold in these spacetimes. However, these spaces still
support ``pseudo-wound" strings -- contractible strings that wrap the
full spatial extent in a given direction. If these strings contract in
time scales that are less than the cosmological Hubble time, they will be
cosmologically irrelevant and cannot support a modified version of the
Brandenberger-Vafa scenario. However, if their contraction time scale
is significantly longer than the Hubble time this raises the
possibility that pseudo-wound modes can play the same role as their
stable, truly wound cousins.

As an illustrative example, consider a compactification involving
an eliptically fibered $K_3$ manifold. $K_3$ has a trivial
fundamental group and yet in the neighborhood of a non-degenerate
fiber the space looks locally like $U \times T^2$ with $U$ an
open set in $CP^1$. A string that winds around a cycle of the
$T^2$ is not topologically stable, but in order to decay it must
move over the base to one of 24 singular fibers on which it can
shrink. If the $CP^1$ base is large or asymmetric in shape (e.g.
the cigar shape discussed in \cite{StringyCosmicStrings}), it is
conceivable that the growth of the toroidal fibers might be
inhibited by the pseudo-wound string modes over cosmologically
interesting timescales.

Similarly, even though toroidal orbifolds generally have trivial
fundamental groups, string configurations that would be topologically
wound on the covering space must find their way to a fixed point of
the group action in order to unwind. This process introduces a time
scale and here we study the lifetime of such pseudo-wound strings on
orbifolds. We employ the same numerical techniques used in toroidal
spacetimes \cite{sak,vil} but adapt them for use on orbifolds. The
$\mathbb Z_4$ symmetry of the basic 2D lattice used in \cite{vil}
allows us to construct several orbifolds which respect the usual 
supersymmetry conditions \cite{orb1}, and we consider:
\[ T^4 / \mathbb Z_2, \ T^4 / \mathbb Z_4, \
T^6 / \mathbb Z_2, \ T^6 / \mathbb Z_4, \ T^6 / \mathbb Z'_4, \ T^6 / (\mathbb 
Z_2 \times \mathbb Z_2)  
\]
where $\mathbb Z'_4$ acts differently than $\mathbb Z_4$, as explained below.  For purposes of comparison, 
we also consider the ``factored," non-supersymmetric
orbifolds
\[ T^2/\mathbb Z_2, \ (T^2/\mathbb Z_2)^2, \ (T^2/\mathbb Z_2)^3,
\ T^2/\mathbb Z_4, \ (T^2/\mathbb Z_4)^2, \ (T^2/\mathbb Z_4)^3. \]
In this way we study the dimensional and topological dependence
of various scenarios.

Our numerical simulations assume that the background spacetime is not
evolving, so while our results shed light on the stability and
persistence of string networks on toroidal orbifolds, they do not
directly describe the dynamical evolution of these spaces. Moreover, 
the Brandenberger-Vafa scenario requires
that the dilaton is evolving freely \cite{tv}, so the spacetime
dynamics cannot be governed by general relativity. If the dilaton is
fixed, the low energy limit of string theory reduces to general
relativity and string networks actually enhance the expansion rate
since they have negative pressure. However, as shown in Appendix A of
Ref.~\cite{tv}, string networks do not induce rapid expansion in
dilaton gravity, and the subspaces where the strings do not annihilate
will expand more slowly than those in which they do.  The mechanism
that fixes the dilaton is not well understood, but it is reasonable to
assume that it is related to the physics which clamps the values of
the moduli that set the size of the internal space, which must happen
if the compactification is to be stable. This leads to a consistent
picture: in addition to a running dilaton, the Brandenberger-Vafa
scenario requires that the moduli be free to evolve if it is to induce
the anisotropy between the three large directions we observe and the
remaining compact directions, and both the moduli and dilaton are
fixed after the Brandenberger-Vafa mechanism has had time to work.

In Section 2 we review the numerical techniques used in the
simulation, including the representation of the orbifolds we study. In
Section 3 we present the results of the simulations and discuss their
significance. In Section 4 we go on to discuss some general features
of string and brane gas models in the context of Calabi-Yau
compactifications, and we identify possible extensions to this work.

\newsection{Numerical simulation}
\subsection{The Smith-Vilenkin Lattice} 

 Our simulations use the lattice model constructed by
Smith and Vilenkin \cite{vil}. The worldsheet gauge is chosen, so 
we must satisfy the conditions 
\begin{eqnarray} \label{gauge1}
{\bf X}' \cdot {\dot {\bf X}} &=& 0, \\ 
\label{gauge2} {\bf
X}{'} {^2} + {\dot {\bf X}} ^2 &=& 1, 
\end{eqnarray} 
as well as the equation of motion 
\begin{equation} 
{\ddot {\bf X}} - {\bf X}'' = 0, 
\end{equation} 
where primes and overdots denote derivatives with respect to the
worldsheet coordinates $\sigma$ and $\tau$.

\begin{figure}[t] \begin{center}
\includegraphics[scale=1.5]{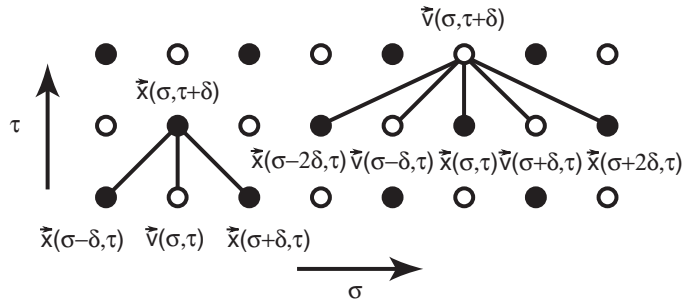} 
\caption{Positions and velocities of the string are evolved based on
nearby positions and velocities from the previous timestep.}
\end{center} 
\end{figure} 

By discretizing the worldsheet coordinates as in
Figure 1, \eq{gauge1} and \eq{gauge2} become 
\begin{eqnarray}
\label{disgauge} {\bf u}' \cdot {\dot {\bf v}} &=& 0, \\ 
{\bf
u}{'} {^2} + {\dot {\bf v}} ^2 &=& 1 \end{eqnarray}
where we have defined
\begin{eqnarray} {\bf u} (\sigma, \tau) &\equiv&
[{\bf X} (\sigma + \delta, \tau) - {\bf X}(\sigma - \delta,
\tau) ]/2 \delta, \\ 
{\bf v} (\sigma, \tau) &\equiv& \{ [{\bf
X} ( \sigma, \tau + \delta) - \frac{1}{2} [ {\bf X} (\sigma +
\delta, \tau) + {\bf X} (\sigma - \delta, \tau)] \} / \delta.
\end{eqnarray} 
The solutions to these equations are
\begin{eqnarray} 
{\bf X} (\sigma, \tau + \delta) &=&
\frac{1}{2} [ {\bf X} (\sigma + \delta, \tau) + {\bf X} (\sigma
- \delta, \tau) ] + {\bf v} (\sigma, \tau) \delta, \\ 
{\bf v}
(\sigma, \tau + \delta) &=& \frac{1}{2} [{\bf v} (\sigma +
\delta, \tau) + {\bf v} (\sigma - \delta, \tau)] \\ 
\nonumber
&+& [ {\bf X} (\sigma + 2 \delta, \tau) - 2 {\bf X} (\sigma,
\tau) + {\bf X} (\sigma - 2 \delta, \tau)]/4 \delta.
\end{eqnarray} 
These are gauge preserving and become exact solutions to the
differential equations \eq{gauge1} and \eq{gauge2} as $\delta
\rightarrow 0$. 

\begin{figure}
\begin{center} \includegraphics[scale=0.6]{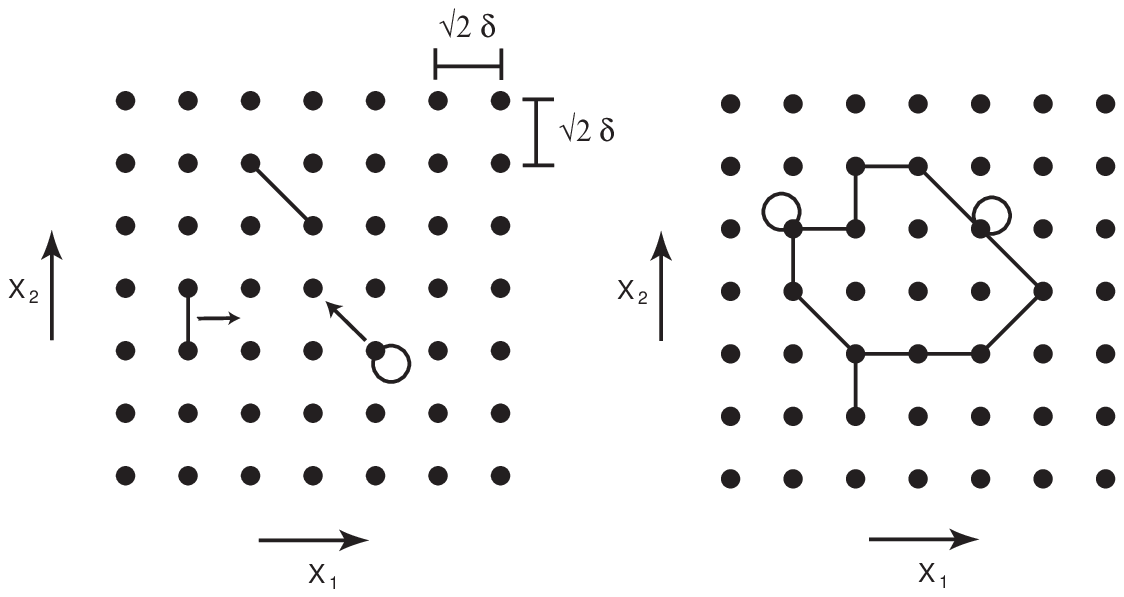}
\caption{(Left) Links in the string are constructed from three
types of segments, moving at a velocity such that the combined
energy (due to velocity and tension) in each is
uniform. (Right) A typical string.} \ \\
\includegraphics[scale=0.6]{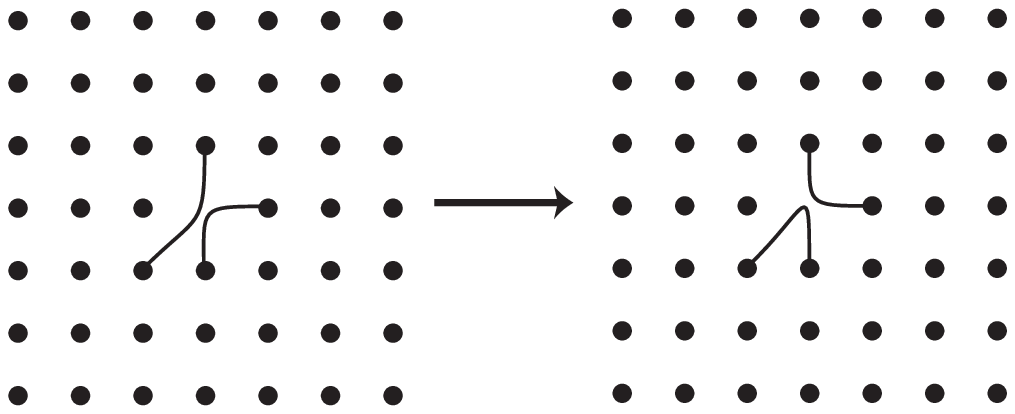} \caption{Interactions
are performed via intercommutation at a single point, which has
been removed for clarity.} 
\end{center} 
\end{figure}

It is essential that the initial conditions respect the gauge. This is
ensured by also discretizing the target space and allowing only three
types of links, as shown in Figure 2 (our target space grid has been
rotated from that of \cite{vil}). The first type has $|{\bf u}| = 1$
and ${\bf v}=0$. Its end points are separated by $|\Delta {\bf X}| = 2
\delta$, which implies that two components of $\Delta {\bf X}$ are
$\pm \sqrt{2} \delta$ and the others are zero. Such links are ``fully
stretched" and are at rest. The second type has $|{\bf u}| = | {\bf v}|
= 1 / \sqrt{2}$, implying that one component of $\Delta {\bf X}$ is
$\pm \sqrt{2} \delta$ and the others are zero, and one component of
${\bf v}$ is $\pm \frac{1}{\sqrt{2}}$ and the others are zero. Such
partially contracted links move at 0.707 light speed perpendicular
to themselves. The third type has ${\bf u} = 0$ and $| {\bf v}| =
1$. Its end points are degenerate, with $\Delta {\bf X}=0$. Two
components of ${\bf v}$ are $\pm \frac{1}{\sqrt{2}}$ and the others
are zero. Such links are fully contracted and move at light speed
diagonal to the coordinate axes. A 2D lattice is shown here but the
generalization to higher dimensions is obvious.

String interactions are achieved via the intercommutation of
two strings which pass through the same lattice site in the target
space, as shown in Figure 3. This preserves the gauge constraint
exactly, as well as energy, momentum and angular momentum. We specify
the coupling constant by assigning a probability $P$ that strings in
such a configuration will in fact intercommute.

\subsection{Simulation of strings on orbifolds}  In Figure 4
we show how $G = \mathbb Z _2$ acts on the manifold $M = T^2$
to produce the pillow-shaped orbifold $\Gamma = T^2/ \mathbb Z
_2$ with fixed points $O$, $P$, $Q$, and $R$. Construction of
$T^2 / \mathbb Z_4$ is similar, producing a space with three
fixed points. Higher-dimensional orbifolds are of course more
difficult to visualize, but are mathematically straightforward.
We construct $\Gamma = T^4 / \mathbb Z_2$, $T^4 / \mathbb
Z_4$, $T^6 / \mathbb Z_2$, and $T^6 / \mathbb Z_4$ by letting $z_i,$ $i=1,
\ldots,D/2$, be the variables on a complex torus defined by 
$z_i \cong z_i + 1 \cong z_i + i$.  The 
group acts on these as follows:
\[ (z_1,z_2) \mapsto (e^{i \theta} z_1, e^{-i \theta} z_2) \]
with $\theta = \pi$ for $G = \mathbb Z_2$ and $\theta = \pi / 2$ 
for $G = \mathbb Z_4$.  For $D=6$, $z_3$ is left invariant under the group 
action.  We construct $T^6 / \mathbb Z'_4$ by acting as
\[ (z_1,z_2,z_3) \mapsto (e^{i \pi /2} z_1, e^{i \pi / 2} z_2, e^{i \pi} z_3) \]
and finally $T^6 / (\mathbb Z_2 \times \mathbb Z_2)$ by
\begin{eqnarray*}
(z_1,z_2,z_3) &\mapsto& (e^{i \pi} z_1, e^{i \pi} z_2, z_3) \\
(z_1,z_2,z_3) &\mapsto& ( z_1, e^{i \pi} z_2, e^{i \pi} z_3).
\end{eqnarray*}
We refer to these as the ``supersymmetric orbifolds".  For the ``factored" 
cases $(T^2/\mathbb Z_2)^{D/2}$ and $(T^2/\mathbb Z_4)^{D/2}$ each $z_i$ is 
acted on by a separate copy of $G$.

\begin{figure} 
\begin{center}
\includegraphics[scale=0.4]{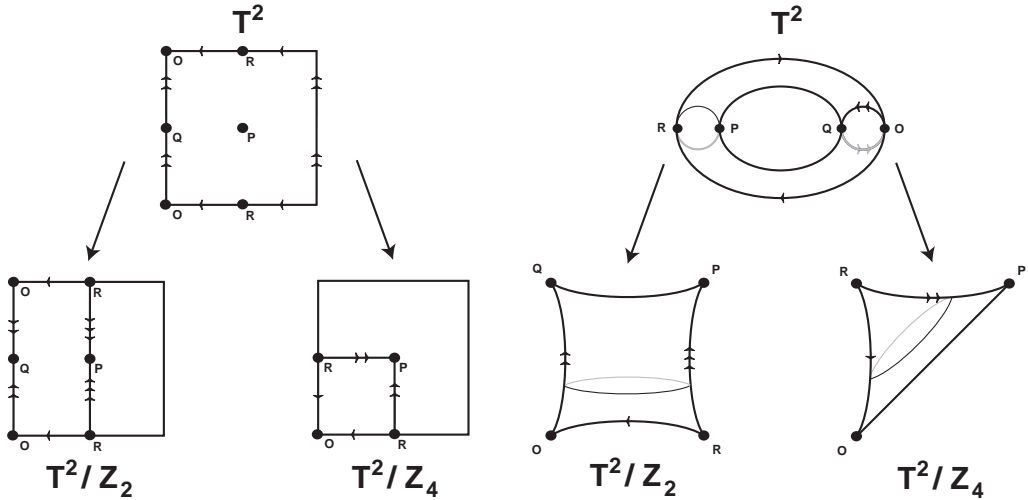} 
\caption{Orbifolding takes
a smooth geometry and creates singularities if the group action has fixed 
points. In these examples,
we take $T^2 \rightarrow T^2 / \mathbb Z_2$ and $T^2
\rightarrow T^2 / \mathbb Z_4$.} 
\end{center} 
\end{figure}

We simulate string evolution on these spaces as follows. When
identifying interactions, the group acts on the position of each
string link, producing an orbit. Identical orbits (rather than
identical positions) then give rise to intercommutation, with the
group action information included in the string's link for future
string evolution.

\section{Simulations and Results}
 We begin by checking our simulation algorithm against previous 
codes, and then perform two sets of computations. In the first we keep the
intercommutation probability at $P=1$, and vary the topology.  In the
second, we focus on topologies which had potential for thermalization
in more than 3+1 dimensions, and study the impact of varying $P$.

\subsection{Critical density computation}
 Previous studies have calculated the critical density in 
string thermodynamics, or the maximum allowed short string density
\cite{sak,vil,sakvil}. We first determine this density as a
computational check. We define a ``long" string in the same way as
previous analyses: a string is ``long" if its extent in at least one
dimension is approximately the length of that dimension. Using a
string network consisting of 100,000 string links and units such that
energy is measured in links and length in $\delta$, we obtain the
following critical densities:
\begin{center}
\begin{tabular}{cc}
$D=3:$ & $0.1715 \pm 0.0083$\\
$D=4:$ & $0.0560 \pm 0.0009$\\
$D=5:$ & $0.0269 \pm 0.0003$\\
\end{tabular}
\end{center}
These agree to within a few percent of the results found in the literature 
\cite{sak,vil,sakvil}.  While the difference exceeds the uncertainty in 
the fit, the results also depend weakly on the overall lattice size, or the 
specific details of the interaction code when more than two strings 
intersect at a point, and these systematic effects are most likely the 
origin of the discrepency.

\subsection{Unwinding rate at constant coupling}
 Our first investigation compares the rates of string unwinding in
 different topologies in the ``strong coupling" regime of $P=1$. This
 allows the most efficient interactions, and hence the greatest
 possible change to some initially wound state.  We chose the
 dimensions so that each space has volume of $\sim (200
\sqrt{2} \delta) ^D$, and a total of 40,000 wound string links. For
example, on $T^2$ we choose $L=200 \sqrt{2} \delta$ and 100 wound
strings, whereas on $T^6 / \mathbb Z_4$ we choose $L=252 \sqrt{2}
\delta$ and 80 wound strings. As shown in Figure 5, the strings are
initially wrapped at random around the dimensions, with half of the
string links being fully extended in a zig-zag pattern and half the
links being fully kinetic with velocity directions chosen at
random. This allows us to construct initial configurations with random
velocities, so that the strings have a fair chance of moving around
and interacting. This kind of winding configuration was one of several proposed in 
\cite{sak}, and suited our purposes best.  The string density was kept low  
so that complications arising from the Hagedorn transition (see
\cite{sakvil,hag,tur}) were avoided and there was no obstacle to all
strings becoming small.

To determine if the result was density-dependent, we then performed
simulations where the lattice length was reduced by factors of 3 and
7, keeping the number of wound strings the same (note that the
number of links in a wound string is proportional to the length of the
dimension).  This significantly increases the density: when the 6D
lengths are reduced by a factor of 3, the density changes by a factor
of $3^5 = 243$.

\begin{figure}[t]
\begin{center}
\includegraphics[scale=0.5]{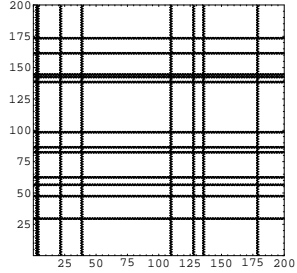}
\caption{Typical initial winding configurations for $T^2$. For clarity we have 
only drawn 20 initially wound strings; in the simulations there were
between 50 and 100.}
\end{center}
\end{figure}

The simulations were run for 20 Hubble times\footnote{We use the term
`Hubble time' to denote the time required for light to propagate
across the size of the universe. This usage is slightly loose since we
are not working with an expanding background.}, which in the lowest
density simulation, was 4000 time steps. In Figure 6 we plot the ratio
of small string energy density $ \rho_{small}$ to the total energy
density $\rho_{total}$ for each configuration. Since all strings began
large, this ratio measures the efficiency with which the strings
interact to form small strings and the speed with which this happens determines
whether the ``pseudo-wound'' strings can have an impact on the overall
cosmological evolution.  For these simulations, we could not use the
``extent" criteria of Refs~\cite{sak,vil,sakvil}, as orbifolds do not
allow simple definitions of length in each dimension. Instead, our
criterion for a string to be ``long" is that the total string length
is at least $L/2$. This is not as descriptive as measuring the
string's extent in each dimension, but since thermodynamics creates a
large gap between ``small" and ``large" strings we believe this serves
our purpose \cite{sak,sakvil}.

Our initial configuration, with half of the links purely kinetic,
allows half of the string to be ``slack". Thus it can lose up to half
of its links and still be wound around the torus or wrap around the
largest cycle of the orbifold.  Since we expect that strings can shed
this slack quickly, the real indication of unwinding is how quickly
the short string density rises after reaching $\rho_{small} /
\rho_{total} \approx 0.5$.
\begin{figure}
\begin{center}
\includegraphics[scale=0.5]{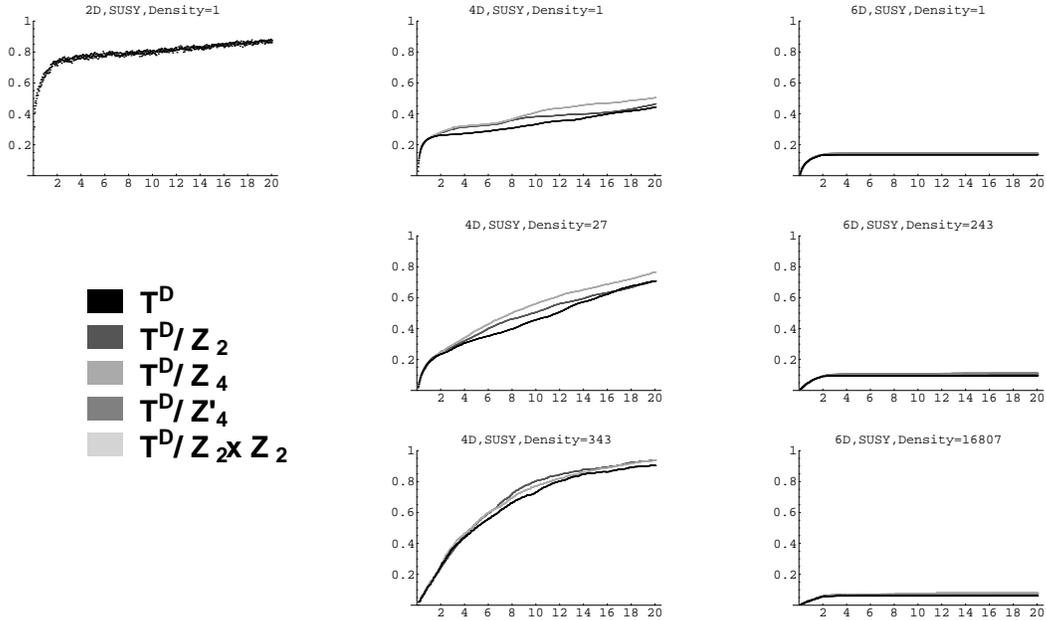}
\caption{Plots of $\rho_{small}/\rho_{total}$ on supersymmetric orbifolds 
as a function of Hubble time for different dimensionalities and densities.  
The 2D cases were only done at the lowest density since we quickly ran 
into Hagedorn issues.}
\end{center}
\end{figure}

The toroidal cases behaved as expected, in agreement with
Ref.~\cite{sak}. With $D=2$ the long strings immediately fragmented to
small strings, whereas with $D=4$ and $D=6$ the long strings
persisted, as expected from naive phase-space arguments based on the
dimensionality of strings.

The results for the orbifold simulations are similar. In all
supersymmetric cases we studied, the unwinding rate was nearly
identical to the toroidal case. This apparent indifference to topology
is presumably due to the high codimension of the fixed point sets,
which makes it difficult for the pseudo-wound strings to find a path
that results in their unwinding. In both of the $4$ and $6$ dimensional
orbifolds we study, the singularities are codimension 4, providing
a significant phase space to support long pseudo-wound strings.

To test this idea, we simulated examples where the topology can be
factored into products of 2-dimensional orbifolds: $(T^2/ \mathbb
Z_2)^{D/2}$ and $(T^2/ \mathbb Z_4)^{D/2}$, yielding lower codimension
fixed point sets.  As explained previously, these were constructed 
analagously to the 
supersymmetric orbifolds, except that each complex torus now has its own 
group action.  In these examples the string can unwind within the
individual 2D orbifolds around which it was initially
wound. Intuitively, we expect these models to produce different
results from the analogous toroidal cases, except for $D=2$ where the
strings will overlap regardless of topology. We emphasize that these
spaces are not realistic candidates for compactification.

\begin{figure} \begin{center}
\includegraphics[scale=0.5]{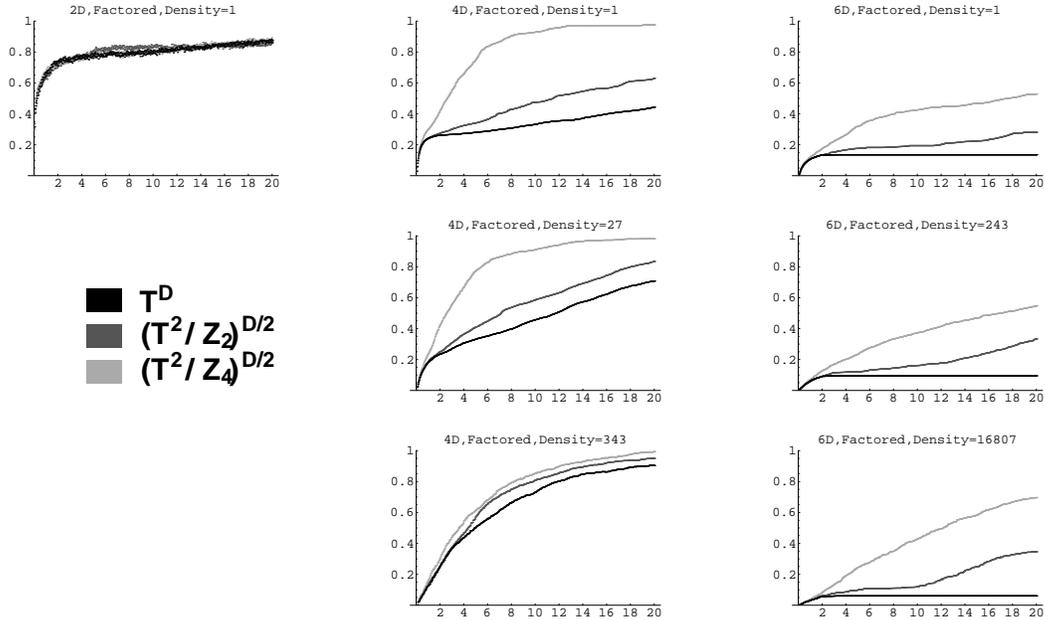} 
\caption{Plots of $\rho_{small}/\rho_{total}$ on factored, non-supersymmetric 
orbifolds as a function of Hubble times for different dimensionalities
and densities.  The 2D cases were only done at the lowest density
since we quickly ran into Hagedorn issues.}
\end{center} \end{figure}

The $T^2/\mathbb Z_2$ case is very similar to the torus, immediately
achieving $\rho_{small}/\rho_{total} \approx 0.5$ and thus unwinding
quickly. For other $(T^2/ \mathbb Z_2)^{D/2}$ spaces, the unwinding is
more efficient than for the torus, but the overall dynamics do not
differ significantly. The $(T^2/\mathbb Z_4)^{D/2}$ cases do differ,
however.  For $D=2$, the story is identical to $T^2$. In $D=4$,
however, the behavior is nearly identical to that of $T^4$ for only
about 1 Hubble time.  After this, the strings continue to contract,
taking only $\sim$ 3 Hubble times to achieve
$\rho_{small}/\rho_{total} \approx 0.5$ and eventually producing
completely short strings. In a cosmological setting, this rapid
unwinding would not be compatible with the Brandenberger-Vafa scheme.
The larger configuration space associated with the $(T^2/\mathbb
Z_4)^3$ prevents rapid string unwinding, although string contraction
is more efficient than that on $T^6$ (see Figure 7).

The intuitive picture of the decay of long strings as a function of
dimensionality is built around two extremes. A low dimensionality has
high string density and thus allows quick but not necessarily complete
fragmentation, due to the Hagedorn transition favoring the creation of
long (and thus likely wound) strings. A high dimensionality has slow
(if any) fragmentation, though once fragmentation takes place there is
no obstacle to all strings remaining small. How the system behaves in
between these extremes depends on the topology. The 2-dimensional
simulations were nearly identical (immediate, but not complete,
unwinding) as were the 6-dimensional cases (slow or nonexistent
unwinding). All 4-dimensional cases begin with moderate unwinding.
After approximately one Hubble time the $T^4$ and $(T^2/\mathbb
Z_2)^2$ fragment very slowly, whereas $(T^2/\mathbb Z_4)^2$ continues
to fragment moderately until all strings are small. This is because
during the first Hubble time fragmentation is only possible due to
mutual interaction. After this, once a string has had enough time to
contract around a fixed point due to its tension, it can
self-interact. This difference is explained by the topology:
$T^2/\mathbb Z_4$ demands that strings wound in the covering space
self-intersect, whereas the others do not.

\subsection{Varied probability of intercommutation}

Of the topologies examined, $(T^2/\mathbb Z_4)^{D/2}$ has the greatest
potential for string unwinding, possibly allowing thermalization in
more than $3+1$ dimensions. Even though this space is not a realistic
candidate for compactification, it allows us to perform one other
useful test. The previous calculations were performed at an
intercommutation probability of 1, but perhaps for sufficiently low
interaction rates the strings would not interact enough to unwind. To
explore this, we repeated the simulation on $(T^2 / \mathbb Z_4)^2$
for up to 5 Hubble times with $0
\leq P \leq 1$, effectively varying the value of the string dilaton,
which determines the intercommutation probability. As shown in Figure
8, for small coupling the unwinding rate sharply rises with $P$ until
$P \sim 0.4$, beyond which it stabilizes. 

\begin{figure}
\begin{center}
\includegraphics[scale=0.45]{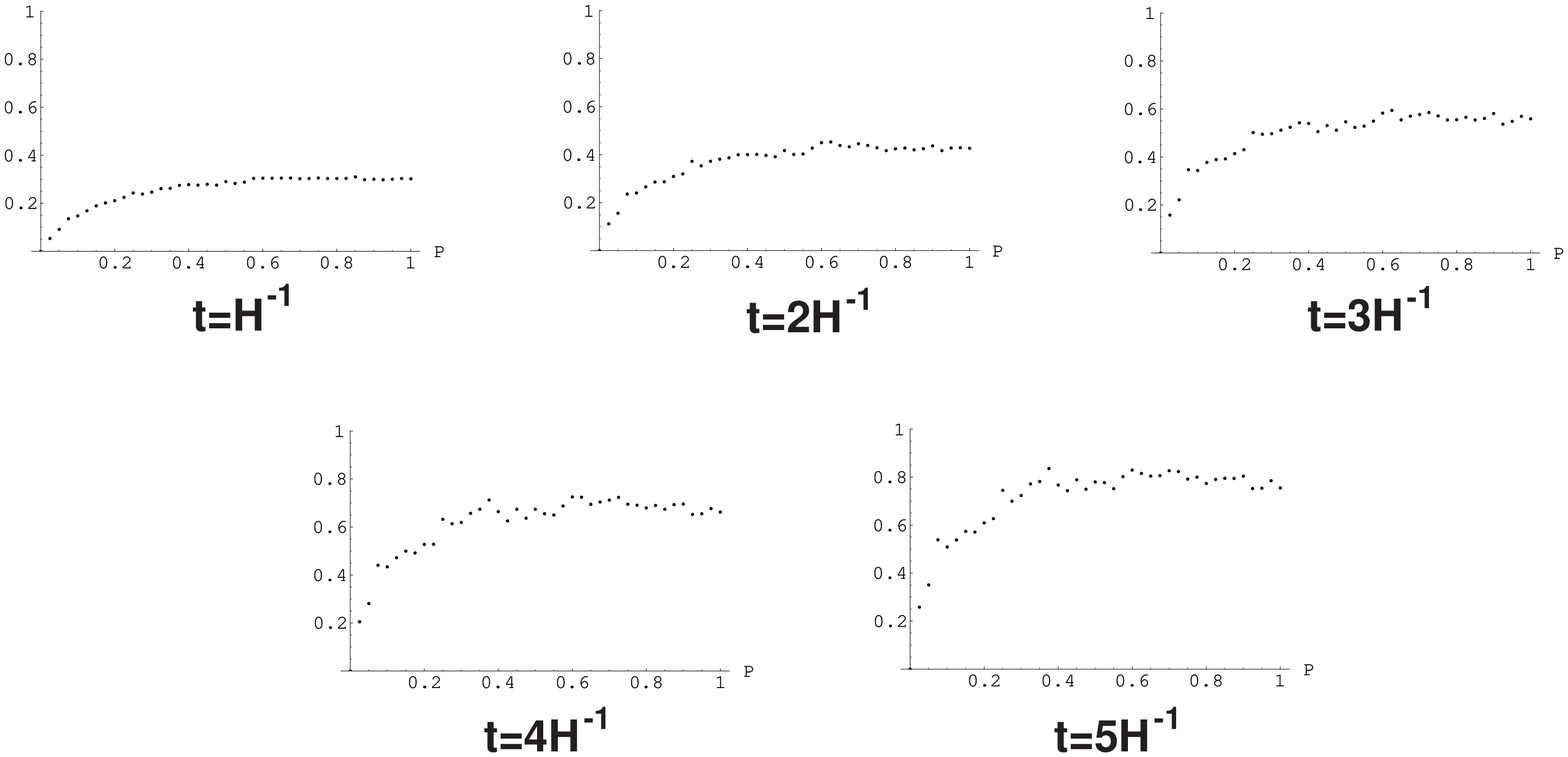}
\caption{Plots of $\rho_{small}/\rho_{total}$ on $(T^2/ \mathbb Z_4)^2$ 
with variable intercommutation 
probabilities $P$ at different time scales.} \ \\
\end{center}
\end{figure}

\section{Branes and General Compactifications}

Our simulations show that pseudo-wound strings on supersymmetry
preserving orbifold compactifications have long
lifetimes. Consequently, they may play an important cosmological role
by inducing anisotropic cosmological expansion similar to that seen
with toroidal compactifications.  So far we have focused solely on
strings, but brane wrappings of higher dimensional cycles can be
important cosmologically, as emphasized in Ref.~\cite{ABE}.  However,
Ref.~\cite{ABE} restricts its analysis to toroidal spaces and it is
worth briefly considering such ideas in more general
compactifications.

In Refs~\cite{bv,ABE}, the authors employ the moduli space
approximation, studying the cosmological evolution of the radial
moduli of the assumed spatial nine-torus topology. Although this
approximation has limited validity in cosmological settings, we also
use it to get a qualitative understanding of the dynamics when branes are
wrapped on more complex compact spaces. Imagine that we compactify
nine spatial dimensions and allow the moduli of the compactification
to be time dependent.  For example, assume that the compactification 
involves the simplest
non-toroidal Calabi-Yau space, namely $K_3 \times T^5$ in type IIA
string theory.\footnote{Here we are specifying the asymmetric topology
by hand and asking if there is a natural mechanism that yields an
asymmetric geometry.}  Just as in \cite{ABE}, the higher dimensional
branes with $p = 8, 6, 4$ generically intersect in nine space
dimensions, and quickly thermalize, leaving 2-branes and 1-branes. We
expect wrapped 2-branes to dominate over wrapped 1-branes in the
energy-momentum tensor, so 2-branes will initially dominate the
cosmology.

The holomorphic 2-cycles in this product space are the 2-cycles on the
$K_3$ and the 2-cycles on the $T^5$, since $K_3$ has no 1-cycles to
hook up with the 1-cycles on the $T^5$. Naive dimension counting then
suggests that thermal fluctuations in the early universe cause some
5-dimensional submanifold to grow large and that efficient
2-brane/anti-2-brane annihilations in this submanifold allow the
geometrical asymmetry to grow.  A natural scenario, independently suggested in 
\cite{k3}, is that the $T^5$ fluctuates to a large size,
leaving the $K_3$ small, and continues to grow ever larger while the
2-branes wrapping the 2-cycles in the $K_3$ keep it small.  However,
this scenario is not unique.  Return to the example in the
Introduction, and imagine that the $K_3$ is eliptically fibered over a
$CP^1$ base and that the initial thermal fluctuation involves $CP^1
\times T^3$. Wrapped 2-branes efficiently annihilate in this subspace
allowing it to continue to grow while 2-branes and 1-branes wrapping
the smaller dimensions do not efficiently annihilate and hence impede
their growth. Within the 5-manifold $CP^1 \times T^3$ we then expect a
thermal fluctuation to drive a larger 3-dimensional subspace within
which wrapped strings can efficiently annihilate. Since $CP^1$ has no
1-cycles, the preferred choice would be a large $CP^1 \times S^1$ so
that wrapped strings keep the remaining $T^2$ small. But if the $T^3$
should fluctuate large first, it is hard to see what would
subsequently stabilize the $CP^1$ at small size.

If, instead, we chose the initial topology to be a Calabi-Yau
three-fold $M$ times, say, a $T^3$, how would things change? At first
sight, one might imagine that the story is essentially the same.  Some
5-dimensional manifold is driven large by a thermal fluctuation, and
within this space a 3-dimensional subspace grows larger still, while
wound 2-branes and 1-branes keep all other dimensions small.  However,
the nontrivial holomorphic cycles in a generic Calabi-Yau space $M$
are isolated.  This means that the annihilation of the branes wrapping
these cycles is enormously more efficient than in the toroidal
cases. For example, if a thermal fluctuation should drive $M$ to grow
momentarily larger than the $T^3$, dimension counting shows that
2-brane annihilation in 6-dimensional space is not efficient and thus
leads us to believe that further expansion would be impeded. However,
if the cycles are isolated, the branes and anti-branes which we assume
to be evenly distributed with no net winding number, immediately
annihilate, allowing the space to continue growing.

A potential way out of this conclusion is to note that
although the holomorphic homology generators are generically isolated,
sufficiently high multiples can be part of continuous families that
sweep through the Calabi-Yau manifold. If branes can wrap these cycles
as independent states, their annihilation would once again be
inefficient, constricting expansion. Even so, a variety of geometrical
evolutions might follow. For example, imagine that $M$ is $K_3$
fibered and that the $K_3$ fibers are themselves elliptically fibered
(over a twisted product of two copies of $CP^1$). Various combinations
of, say, the $CP^1$s in the base and the $T^3$ or its submanifolds
could receive the thermal fluctuation driving a large 5-dimensional
submanifold while wrapped branes keep all other dimensions small. For
instance, if $CP^1 \times T^3$ gets large we are in the same situation
as in the last paragraph.

In some of the above toy scenarios, the geometrical route we
illustrated involved including part of the Calabi-Yau space or the $K_3$
into the large spacetime manifold, partially erasing the motivation
for the initial choice of vacuum topology. If the Calabi-Yau space is
multiply connected or if, using the results of the last section, it is
simply connected but has a sufficiently robust spectrum of
pseudo-wound strings, then we can imagine a scenario in which the
Calabi-Yau stays small.  Namely, in $M \times T^3$ if a
two-dimensional subspace of $M$ together with the $T^3$ should
initially become larger than the other dimensions and if the $T^3$
then receives an additional thermal fluctuation to large size while
wound or pseudo-wound strings keep the somewhat asymmetric Calabi-Yau
space small, we would evolve into the more familiar situation of a small
Calabi-Yau space for the extra dimensions. Moreover, in the moduli
space approximation, some Calabi-Yau spaces simply do not have the
necessary flexibility for all independent subspaces to get large
(e.g. examples with one Kahler moduli) and hence we would then expect
a thermal fluctuation to make either the $T^3$ large from the outset,
or the whole of $M$ large from the outset. If wound or pseudo-wound
strings can impede the growth of $M$, this would favor $T^3$ getting
large, returning us to the usual scenario of a small Calabi-Yau space.

\section{Conclusion}

We have studied the unwinding time scale for pseudo-wound strings on
toroidal orbifolds to decay to short string states and found that in
cases relevant for supersymmetric string compactifications the time
scales can be many times the Hubble time. This suggests that such
strings may play an important cosmological role, a prospect that is
worthy of further study.  Ideally, one would tackle this problem in a
dynamical spacetime, fully incorporating the general relativistic
backreaction of the strings on the geometry.  This would then permit
us to take up the question of geometrical homogeneity within the
compact spatial dimensions. Additionally, while some work has been
done on extending the Brandenberger-Vafa mechanism from string gases
to brane gases, there is much more to be done. As we briefly mentioned,
Calabi-Yau manifolds have numerous non-trivial cycles around which
branes can wrap but since these cycles are often isolated, the
subsequent dynamics can be quite different from what one expects on
the torus and it would be interesting to determine the resulting
cosmological implications.  Finally, while this scenario can produce an 
initial anisotropy between the large and small directions, it does not 
guarantee that the compactification is stable at late times, and this 
problem needs to be addressed separately.

\section*{Acknowledgments}
 We are grateful to R. Brandenberger, D. Easson and
A. Vilenkin for useful discussions. The work of R.E. is supported by
the Columbia University Academic Quality Fund and the Department of
Energy. The work of B.R.G. is supported in part by the DOE grant
DE-FG02-92ER40699B. M.G.J. is supported by a GAANN Fellowship from the
United States Department of Education and by a Pfister
Fellowship. ISCAP gratefully acknowledges the generous support of
Ohrstrom Foundation.

\end{document}